\documentclass[aps,amsfonts,pre,twocolumn,superscriptaddress,showpacs]{revtex4-1}
\usepackage{epsfig,amsmath,amssymb,bm,epsf,graphics,psfrag,verbatim,subfigure}

\usepackage[normalem]{ulem}
\usepackage{float}
\usepackage[usenames]{color}
\usepackage{graphicx}
\usepackage{ulem}
\usepackage{amsfonts}
\usepackage{amsmath}
\usepackage{natbib}
\usepackage{epsfig,amsmath,amssymb,bm,epsf,graphics,psfrag,verbatim,subfigure}

\newcommand{\TrQ}{ {\rm Tr}\  \mathbf{Q} }
\def\tmu{\bar{\mu}}
\def\tv{\tilde{\mathbf{v}}}

\begin{document}

\title{Alignment and Nonlinear Elasticity in Biopolymer Gels}
\author{Jingchen Feng}
\affiliation{Bioengineering Department and Center for Theoretical Biological Physics, \\Rice University, Houston TX, 77251-1892}
\author{Herbert Levine} 
\affiliation{Bioengineering Department and Center for Theoretical Biological Physics, \\Rice University, Houston TX, 77251-1892}
\author{Xiaoming Mao}
\affiliation{Department of Physics, University of Michigan, Ann Arbor MI 48109-1040}
\author{Leonard M. Sander}
\affiliation{Physics \& Complex Systems, University of Michigan, Ann Arbor MI 48109-1040}
\date{\today}
\begin{abstract}
We present a Landau type theory for the non-linear elasticity of biopolymer gels with a part of the order parameter describing induced nematic order {of fibers in the gel}. We attribute  the non-linear elastic behavior of these materials 
to {fiber alignment induced by strain.}
We suggest an application to contact guidance of cell motility in tissue. We compare our theory to simulation of a disordered lattice model for biopolymers. We treat homogeneous deformations such as {simple shear, hydrostatic expansion, and simple extension, and obtain good agreement between theory and simulation. }
We also  consider a localized perturbation which is  a simple model for a contracting cell in a medium.

\end{abstract}

\pacs{62.20.D-, 87.10.Pq, 87.17.Jj}

\maketitle

Bio-polymer gels are complicated materials consisting of {a network of} cross-linked polymer chains. Almost all of these materials show non-linear elasticity, usually strain-stiffening. Typically, the shear modulus increases by an order of magnitude under applied strain~\cite{Wang95,Gardel04,Storm2005}.  These materials are important in biology; for example, in the body there is almost always a gel, the extracellular matrix (ECM) which gives tissue its structure. The biopolymer Collagen-I is the most common constituent of the ECM. When cells  migrate within tissue -- for example  in wound healing and cancer invasion --they crawl  by attaching to the gel fibers \cite{Sander13}. 
When cells move in this fashion they deform the ECM. However, deformation also affects cell motility. In particular, if the fibers in the ECM are partially aligned, cells tend to move along the aligned  direction -- this is called \emph{contact guidance} \cite{AgudeloGarcia11,Vader09}. Since cells deform and align the ECM, and their motility is affected by this deformation, mechanically mediated  \emph{cell-cell interactions} are to be expected~\cite{Vader09,Sander13}. 
In this paper we formulate a theory for the non-linear elasticity in biopolymers to give a framework  for understanding  moving cells. 

We concentrate on a model for Collagen-I. We note that it is an athermal biopolymer: i.e. the elasticity of the individual fibers is purely mechanical because the thermal correlation length is far larger than the mesh size \cite{Stein10}. The model we present is specialized to this case. Further, the strain on individual fibers is modest so that the non-linear strain-stiffening of the matrix arises from linear elements \cite{Onck05,Stein10} in contrast to other biopolymers \cite{Storm2005}. For small strains the elasticity is dominated by the (small) bending modulus {of the fibers} -- different parts of the disordered material turn with respect to one another so that the deformation is non-affine because there are soft bending modes. As strains increase,  bending modes are ``exhausted" and for large strains the material is aligned, and must stretch (Fig.~\ref{FIG:bending}).  {The large stretching modulus then determines the response and the deformation is affine}. A plausible order parameter to describe the transition is precisely the \emph{alignment of the fibers}: it is a  measure of the exhaustion of the soft modes. The claim is that the 
 non-linear elasticity of Collagen-I is usefully connected to  the  ``hidden" variable, alignment.

To quantify  alignment  we use the nematic order parameter, familiar from the study of liquid crystals. Collagen-I does not have a nematic instability, but it can be aligned by stress to give a non-zero value of 
the nematic tensor, $\mathbf{Q}$, describing the local directions of the polymers,
\begin{align}
	\mathbf{Q}(\mathbf{r})\equiv \langle \hat{\nu}\hat{\nu} - \frac{1}{d}I \rangle.
\end{align}
Here $\hat{\nu}$ is the unit vector pointing along the orientation of the polymers within the volume element at $\mathbf{r}$, and the average $\langle \cdots \rangle$ is over all fibers within the volume element, weighted by  the fiber  length. In what follows we will use a scalar, $q, \ 0 \le q \le1$, to characterize the strength of the alignment. We define 
$q = [{d}/{(d-1)}]\lambda_m,$
where $\lambda_m$ is the largest eigenvalue of $\mathbf{Q}$. In two dimensions, $q=\langle \cos(2\theta) \rangle$, where $\theta$ is the direction of the polymer chain {measured from the aligned direction.  This measure was used in the experiments of  Ref.~\cite{Vader09}.  

We proceed by  formulating a Landau theory for biopolymers along the lines suggested in Ref.~\cite{Lubensky02} for nematic elastomers, We use  two coupled order parameters, the strain tensor and $\mathbf{Q}$. We test the framework by fitting our theory to a disordered lattice model which share many feature with Collagen-I \cite{Das2007,Broedersz11,Das12,Mao13}. 
The unique feature of our approach is that \emph{all of the Landau coefficients are determined explicitly}, allowing a detailed test of the modeling.
Then we  insert a ``cell", i.e., a localized source of deformation, to begin to address the questions  above. 

First, we model the nonlinear elastic energy under shear deformation only. Volume-changing deformations will be treated later.  
The energy is a functional of the deformation gradient tensor \cite{Lubensky02}:
$\Lambda_{ij} = \partial R_i/\partial r_j= \delta_{ij} + \partial_j u_i$
Here $\mathbf{r}$ is a point in the reference space, $\mathbf{R}$ is its image in the deformed space {and $i,j$ are Cartesian indices}. From this we form the left Cauchy-Green strain tensor:
\begin{equation}
\mathbf{v}= (1/2) (\Lambda \Lambda^T - \mathbb{I}).
\end{equation}
The volume element in the deformed space is ${\rm det}(\Lambda) \to 1 + {\rm Tr}\ \mathbf{v}$, {where the last form is is for the linear regime.}

For the moment we consider simple shear so that, to lowest order, we need the traceless part of $\mathbf{v}$ which we call $\mathbf{\tilde{v}}$. We propose the  following  free energy:
\begin{align}
\label{EQ:FE}
	F_s = \int \Big\lbrack 
	 \tmu \textrm{Tr} \ \mathbf{\tilde{v}}^2 
	-t  \textrm{Tr} (\mathbf{\tilde{v}}\cdot \mathbf{Q}) + V(\mathbf{Q}) \Big\rbrack d^d r ,
\end{align}
where  $\tmu$ is the shear modulus. The bar on $\tmu$ means that it is a ``bare" value. The observed shear modulus, $\mu$, is gotten by renormalizing $\tmu$ by coupling to $\mathbf{Q}$, as we will see.     The potential for the nematic order, $V(\mathbf{Q})$, is the energy cost to align the network against the constraints.

The  term  $-t  \textrm{Tr} (\mathbf{\tilde{v}}\cdot \mathbf{Q})$  is the leading order coupling between strain and alignment allowed  by symmetry. The order parameter, $\mathbf{Q}$, which describes the alignment induced by strain, transforms as a tensor in the deformed state so that it has the same symmetry as $\mathbf{\tilde{v}}$. Their contraction is a scalar  that can enter the free energy.  Because $\mathbf{Q}$ is traceless, in leading order the alignment responds to shear  and not to hydrostatic deformations.  {This form of coupling between strain and local material anisotropy is similar to  Cosserat (micropolar) elasticity~\cite{Cosserat1909,Kunin1983} }

The auxiliary variable $\mathbf{Q}$ is not necessary in our formulation. We could, in traditional fashion, write an explicitly non-linear theory instead of Eq. (\ref{EQ:FE}) -- it would correspond to our model  after ``integrating out" $\mathbf{Q}$. We do not do this for several reasons: first, $\mathbf{Q}$ is observable \cite{Vader09} and gives useful information about the system. Also, if we formulate the model in terms of $\mathbf{Q}$ it is rather simple and physically motivated. There is no obvious way to pick the non-linear terms in a traditional formulation, but they arise naturally in our method. Finally, this theory gives a good way to deal with contact guidance.

For small deformations, linear elasticity, we keep the leading order term for the potential, $V \approx (A/2) \mathrm{Tr} \mathbf{Q}^2 $. 
Minimizing $F$ for fixed $\mathbf{v}$ gives
	$\mathbf{Q} = (t/A)  \tilde{\mathbf{v} }$.
Thus \emph{strain induced alignment} is a linear response for small deformations.  Since this relation is determined by symmetry, we expect it to hold for all  biopolymer gels.  Now $\mathbf{Q}$ can  be eliminated from $F$, recovering linear elasticity:
\begin{align}
	F \to \int \Big\lbrack 
	\mu \textrm{Tr} \mathbf{\tilde{v}}^2 
	\Big\rbrack d^d r; \quad \mu = \tmu - {t^2}/{2A}
	\end{align}
with an effective shear modulus,
	$\mu < \tmu$. 

At larger deformations, nonlinear terms in $V$ start to dominate and $\mathbf{Q}$ falls below the linear response value.   The effective shear modulus increases from its value in the linear regime, giving strain stiffening.  In the extreme case of $\mathbf{Q}$ reaching a saturated value $\mathbf{Q}_{\textrm{max}}$
independent of $\mathbf{v}$, we get $\tmu \simeq \mu$.  

The nonlinearity in  $V(\mathbf{Q})$ can be interpreted.  At small strain  the deformation in a {dilute} network consists mostly of {nonaffine deformations involving} bending of fibers. At greater strain, the deformation {crosses over into deformation involving} stretching of fibers, because the easy bending modes 
{become} exhausted, 
{as shown in Fig. \ref{FIG:bending}.  As a result, the increase of $\mathbf{Q}$  will eventually saturate.} In contrast, for {denser} networks, there are fewer weak modes, even for small deformations the system is 
{mostly affine,} stretching dominated, 
{and the alignment  follows the geometry of the deformation}.  
{As a result, the crossover disappears, and the network has a more linear increase of $\mathbf{Q}$ as strain increases.}  We  {thus} expect the nonlinearity in  $V(\mathbf{Q})$ to be strong for dilute networks and weak for dense ones.  This is verified in simulations,  see below.
\begin{figure}
	\centering
	
		\includegraphics[width=.4\textwidth]{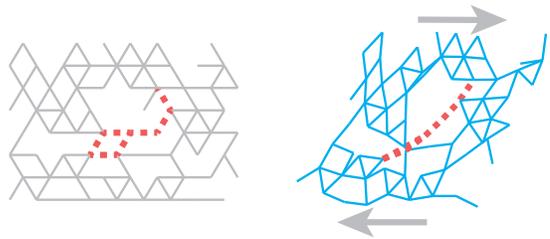}
	\caption{ {A small area of a network before (left) and after (right) simple shear. Dashed lines: the part of the fiber segments that align with the direction of strain after easy bending modes are exhausted.  Solid lines: other fiber segments.}}
	\label{FIG:bending}
\end{figure}

Now we consider volume-changing deformations such as hydrostatic expansion. There is a conceptual difficulty here since, for this case, there is no average alignment. Nevertheless, there can be non-linearity in the bulk modulus.  {For example, the filamentous triangular lattice model~\cite{Das2007,Broedersz11,Das12,Mao13} (discussed in more detail below) shows similar strain-stiffening effects in bulk and shear moduli (Fig.~\ref{FIG:ShearAll}).  Such strain-stiffening in hydrostatic expansion has been studied in Ref.~ \cite{Sheinman12} via simulation and in effective medium theory.  To capture this effect in our Landau-type theory, we note }
that  even for hydrostatic expansion, there will be \emph{local} alignment and weak parts of the disordered system will rotate and the deformation is non-affine.  {At large strain, deformations cross over to affine, stretching dominant, similar to the case of shear.}

We  treat such effects of disorder by introducing a random tensor field, $\mathbf{h}(\mathbf{r})$ with $\langle \mathbf{h} \rangle=0, \left<\mathbf{h}(\mathbf{r})\cdot \mathbf{h}(\mathbf{s})\right> = g\delta(\mathbf{r}-\mathbf{s})\mathbb{I}$ which accounts for  {disorder effects on the alignment field.  Using $\mathbf{h}$ we can form a scalar, ${\rm Tr}(\mathbf{h}\cdot \mathbf{Q})$, which can couple to volume change. }
To characterize the volume change  we introduce $Y = 2(\sqrt{{\rm det}\Lambda} -1) \to {\rm Tr}\mathbf{v}.$  The last form is in the linear regime. We have specialized to two dimensions so that we can compare to a two-dimensional model below.  {This form of volume change is proportional to the extension of the fibers in the network.  The elastic energy for affine deformations is strictly quadratic in $Y$.}
Now we add to Eq.~(\ref{EQ:FE}) the following terms:
\begin{equation}
\label{EQ:FE2}
F_h = \int \Big\lbrack 
	 \frac{1}{2} \bar{K}Y^2 -Y{\rm Tr}(\mathbf{h}\cdot \mathbf{Q})   \Big\rbrack d^d r.
\end{equation}
 {The coupling term between $Y$ and ${\rm Tr}(\mathbf{h}\cdot \mathbf{Q}) $ is the leading order term that controls the nonlinearity in the bulk modulus.}  (There is no term of the form ${\rm Tr}(\mathbf{h}\cdot \mathbf{v})$ because it would give a net stress on the lattice in the reference state.)
The total energy is $F_s + F_h$. 
Once more, we have a bare quantity, $\bar{K}$, the bare bulk modulus. 

We follow the steps above to see how renormalization of the bulk modulus occurs. For a hydrostatic expansion, we minimize $F$ with respect to $\mathbf{Q}$ and find, in the linear regime, $\mathbf{Q}= ({\rm Tr} \mathbf{v}/A)\mathbf{h}$, a linear response. Putting this form back into $F$ and taking the disorder average gives, as in the case of $\bar{\mu}$:
$K = \bar{K} - 2g/A.$
For large strain the renormalization is smaller, and $K \to \bar{K}$, as above.

To calibrate Eqs. (\ref{EQ:FE}, \ref{EQ:FE2}) we apply the theory  to a  model for bio-polymer gels based on a triangular lattice of lattice constant $a$, in which each bond is present with a probability $p$ {~\cite{Das2007,Broedersz11,Das12,Mao13}}.  Straight lines in this lattice, which have average length $(1-p)^{-1}$, are identified as fibers with stretching stiffness $k$ and bending stiffness $\kappa$.  The lattice sites are freely rotating crosslinks.  The Hamiltonian is:
\begin{align}
	E = \frac{k}{2a} \sum_{\langle ij\rangle} g_{ij} (\vert \mathbf{R}_{ij}\vert -a)^2
	+ \frac{\kappa}{2a} \sum_{\langle ijk\rangle} g_{ij} g_{jk} \Delta \theta_{ijk}^2 ,
\end{align}
where $g_{ij}=1$ for bonds that are present, and $0$ for removed ones.  The first term is the stretching energy; $\vert \mathbf{R}_{ij}\vert$ is the distance between sites $i$ and $j$ in the deformed state. The second term is bending;  $\langle ijk\rangle$ labels three consecutive sites along a straight line in the reference state, and $\Delta \theta_{ijk}$ the change of angle determined by them.

The linear elasticity of this model is largely controlled by the central force isostatic point at $p_c\simeq 2/3$.  For $p> p_c$ the deformations are mostly affine; below $p_c$ disorder induced by the removal of bonds leads to non-affine response.
This is because the bending stiffness of the fibers is much smaller than the  stretching stiffness: $\kappa/(ka^2) \ll 1$.  When $\kappa=0$, the system is a central-force lattice, with a rigidity percolation transition at $p_c$.  The network with weak bending can be viewed as central-force network with a relevant perturbation of bending stiffness. The elasticity of such networks can be thought of as a crossover at the central-force isostatic point \cite{Broedersz11}.

We  studied this model  in the nonlinear regime using 128$\times$128 lattices. We applied three types of homogeneous deformations: simple shear, hydrostatic expansion, and simple extension. We found the stress and $q$ as functions of strain, $\gamma$, at various values of $p$ and $\kappa/(ka^2)$.  

For shear, some results are shown in Fig.~\ref{FIG:ShearAll}.  In agreement with our theory, the nematic order tensor $\mathbf{Q}$ does have the form  $\eta(\gamma) \tv$ where $\eta(\gamma)$ is a scalar.  Thus the orientation of the nematic order is determined. The strength of the alignment, $q = \langle \cos 2\theta \rangle$,  is a non-trivial function of $\gamma$. For small strain, the alignment is less than the purely geometric effect which occurs for affine deformations because of bending; see \cite{SM}. The strain at which $q$ falls below a linear dependence on $\gamma$ is hard to see in Fig.~\ref{FIG:ShearAll}(b), but it is not inconsistent with the onset of non-linearity in Fig.~\ref{FIG:ShearAll}(a). For very large $\gamma$, $q \to 1$, but this is  beyond the range of validity of our theory.

  \begin{figure}
	\centering
		\subfigure[]{\includegraphics[width=.23\textwidth]{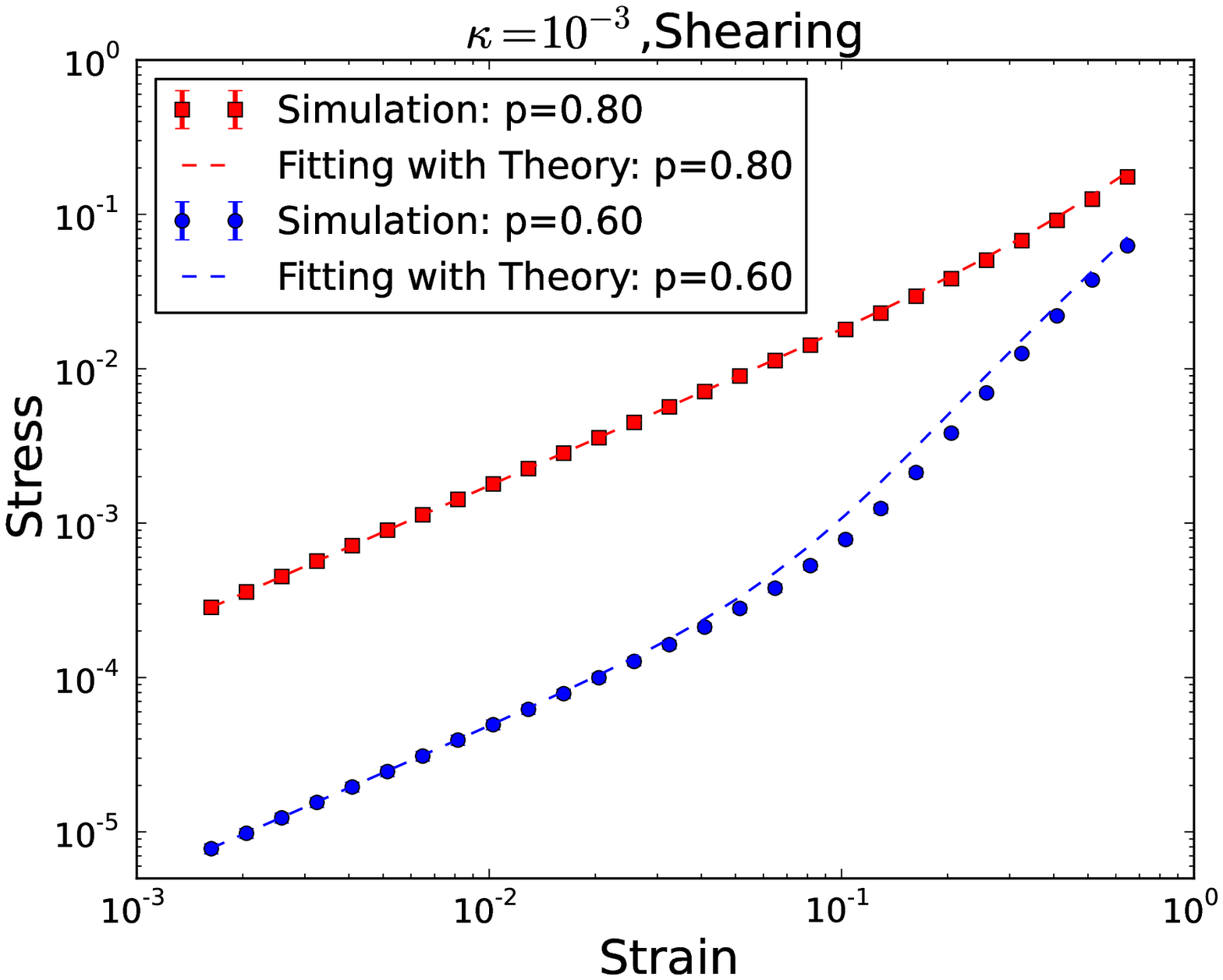}}
		\subfigure[]{\includegraphics[width=.23\textwidth]{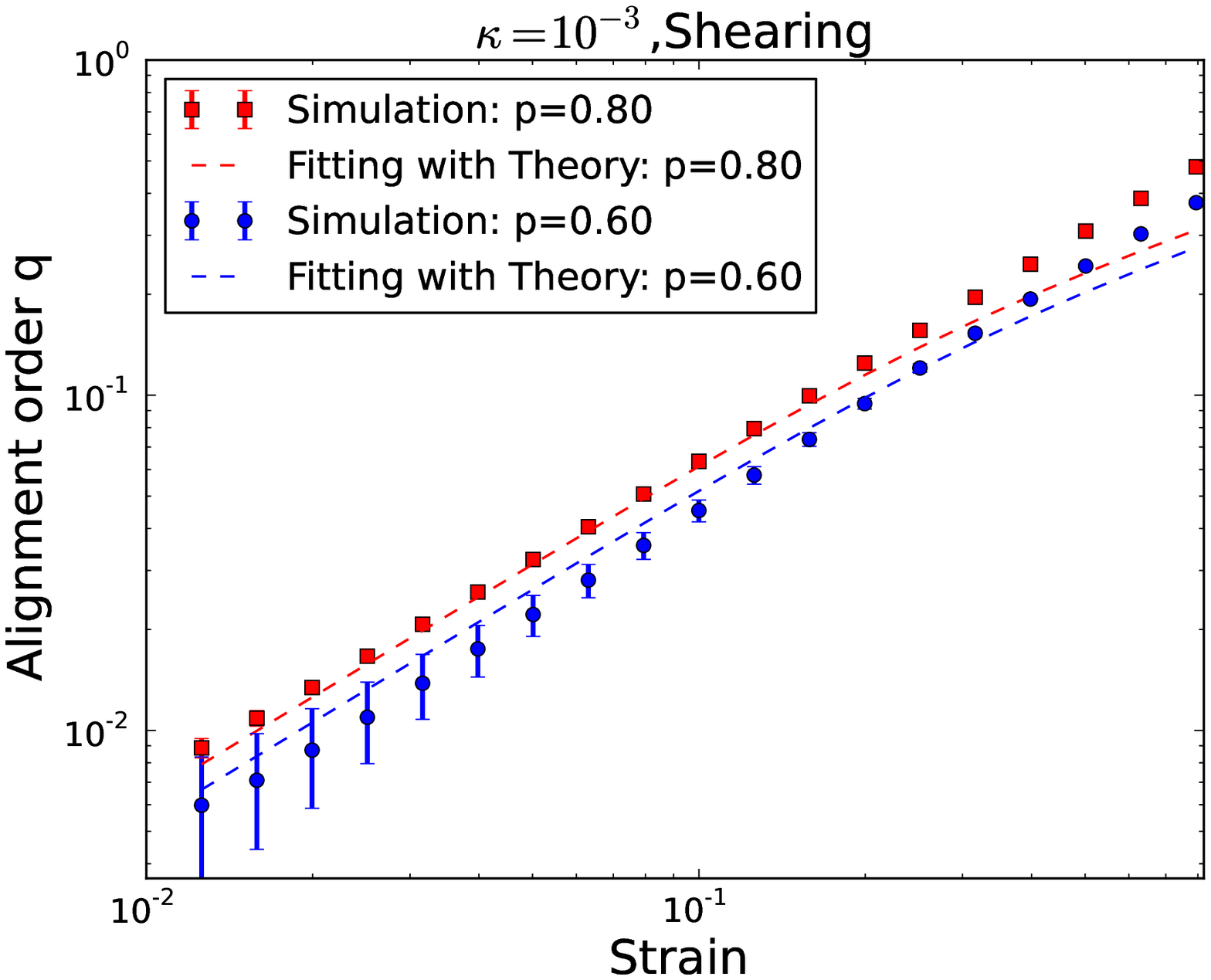}}
		\subfigure[]{\includegraphics[width=.23\textwidth]{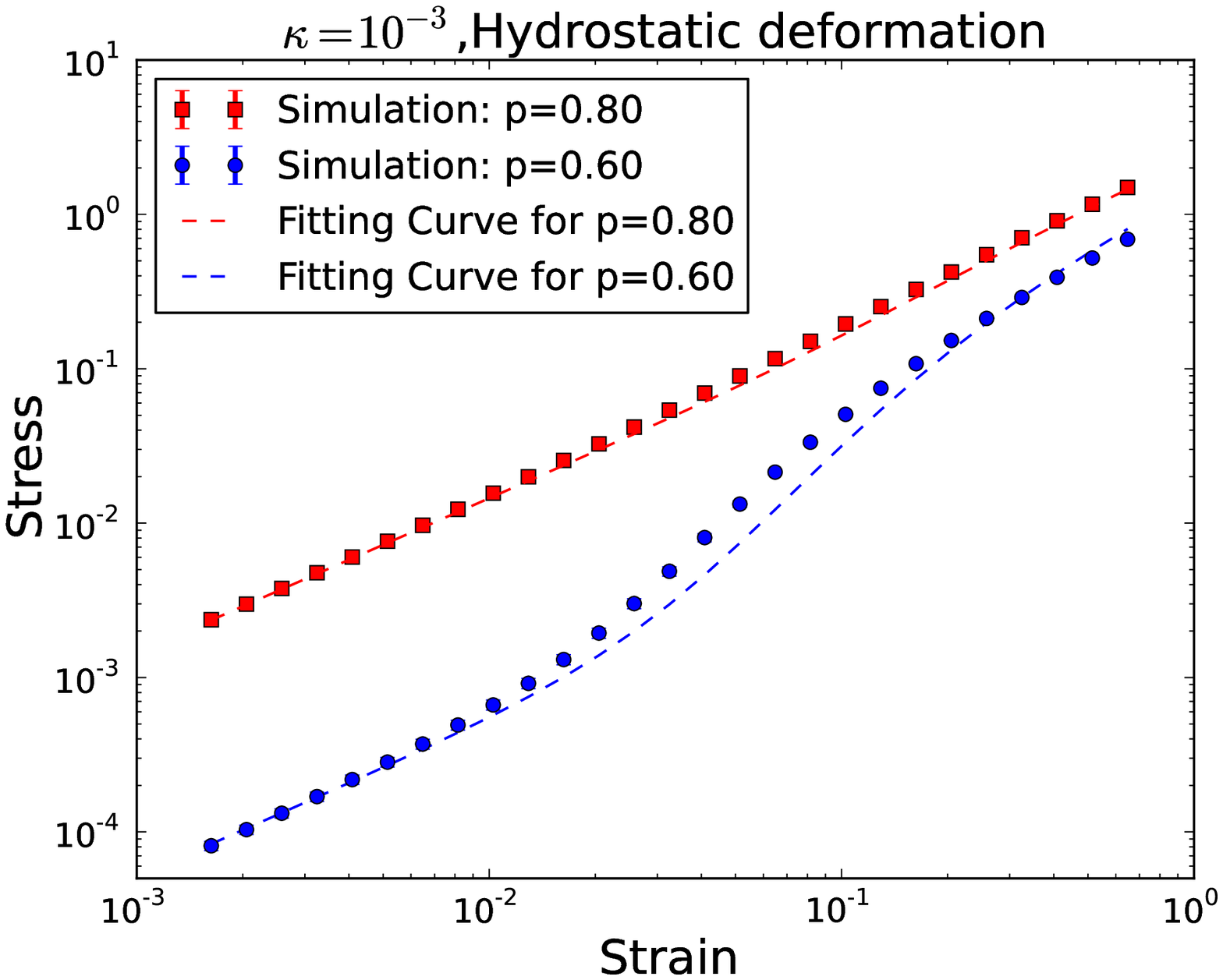}}
		\subfigure[]{\includegraphics[width=.23\textwidth]{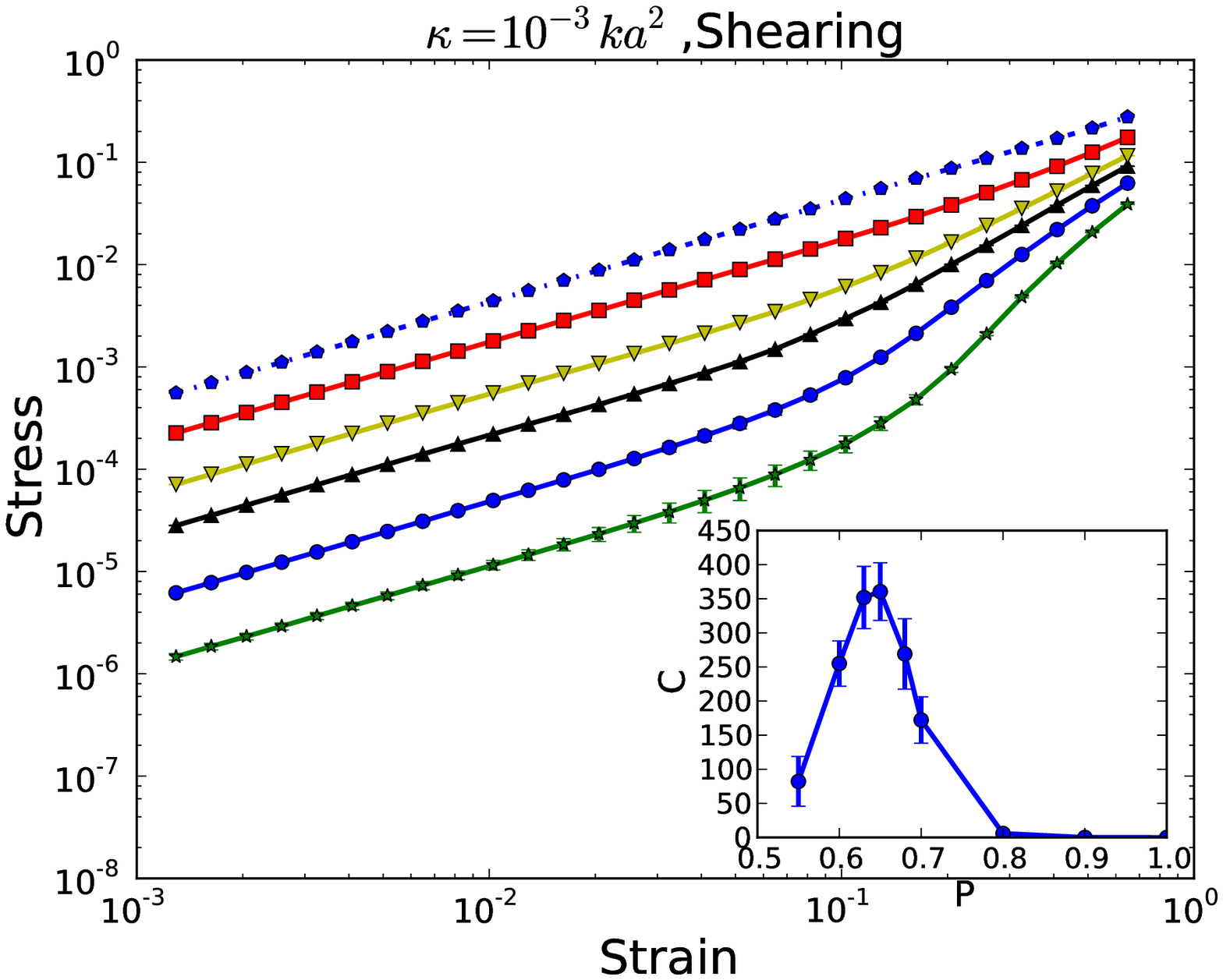}}
	\caption{ {Simulation (data points) and fit to our model (dashed curves) based on Eqs.~(\ref{EQ:FE}, \ref{EQ:FE2}), for lattices with $p=0.6, 0.8$, $\kappa/(ka^2)=10^{-3}$. (a) Stress-strain, simple shear. (b) Nematic-strain, simple shear.  (c) Stress-strain, hydrostatic expansion.  (d) Behavior with $p$: stress-strain, for various $p$ (From top to bottom, p values are 1.00, 0.80, 0.70, 0.65, 0.60, 0.55). Inset, fit parameter $C(p)$, which controls the nonlinearity in $V$.}}
	\label{FIG:ShearAll}
\end{figure}  

These results are consistent our picture. Shear-stiffening is strong for $p \ll p_c $ and vanishes for $p>p_c$. The characteristic strain, $\gamma^*$, where the strain-stiffening takes place is large at small $p$ and  vanishes near $p_c$.  These observations are consistent  the exhaustion of  bending modes as the network enters the stretching dominated regime.   They are related to  stiffening in jammed packings, \cite{Wyart2008a}, where scaling $\gamma^* \sim \vert p-p_c\vert$ occurs.  

To make the comparison with our lattice simulation quantitative, we use Eqs. (\ref{EQ:FE}, \ref{EQ:FE2}) in two dimensions, and expand the potential $V$, up to fourth order in $\mathbf{Q}$:
\begin{align}
	V(\mathbf{Q}) = ({A}/{2})\TrQ^2 + ({C}/{4!}) \TrQ^4 
\end{align}
The odd  terms vanish by symmetry in two dimensions.  Since we are using an expansion of this type, we should not expect the scheme to work deep in the non-linear regime; we expect quantitative results for relatively small $q$. The parameters in the theory are $\{\bar{K}, \tmu, g, t, A, C\}$.  

Consider the linear regime, which is characterized by three independent slopes  for stress-strain curves in simple shear and in hydrostatic expansion, and the nematic-strain curve in simple shear; the corresponding slopes are the shear modulus, $\mu = \tmu - t^2/(2A)$, the bulk modulus, $K=\bar{K} - 2g/A$, and $t/A$. By symmetry, there cannot be any average induced nematic order in the hydrostatic expansion case, and this is what we find  {in simulation}.
The rest of the parameters are gotten in the nonlinear regime.  We  fit them using simulation for hydrostatic expansion for $g$ and  simple shear for the rest, separately via least-square fitting.  The details of the computation are given in \cite{SM}. The results are shown in Figure~\ref{FIG:ShearAll}. 

Strain-stiffening depends strongly on $p$.  This is captured by the parameter $C$, which controls where the potential  becomes strongly nonlinear and increases $\mu$.  Consider $C(p)$,  Fig.~\ref{FIG:ShearAll}c: There is a sharp peak at $p_c$, showing that $\gamma^{*}$ vanishes as $p\to p_c^{-}$. Also, consider the strain-alignment curve. Its slope is  $t/A$ for small strain. When $p \to 1$ the deformations are affine, the alignment is a geometric quantity so that $t/A \to $ a constant. This is consistent with our fitting results. However, above $p_c$ we expect little renormalization of $\mu$, so that $t^2/A$ is small. We conclude, and we do find that both $t$ and $A$ go to zero above $p_c$ in such a way that their ratio is constant.

We used the fitted parameters to calculate the stress-strain and nematic-strain curves for simple extension. In  Fig.~\ref{FIG:Extension} we compare to  simulation.  The agreement is fairly good, except for $q$ at large alignments {which is beyond the range of validity of our theory}.
\begin{figure}
	\centering
		\subfigure[]{\includegraphics[width=.23\textwidth]{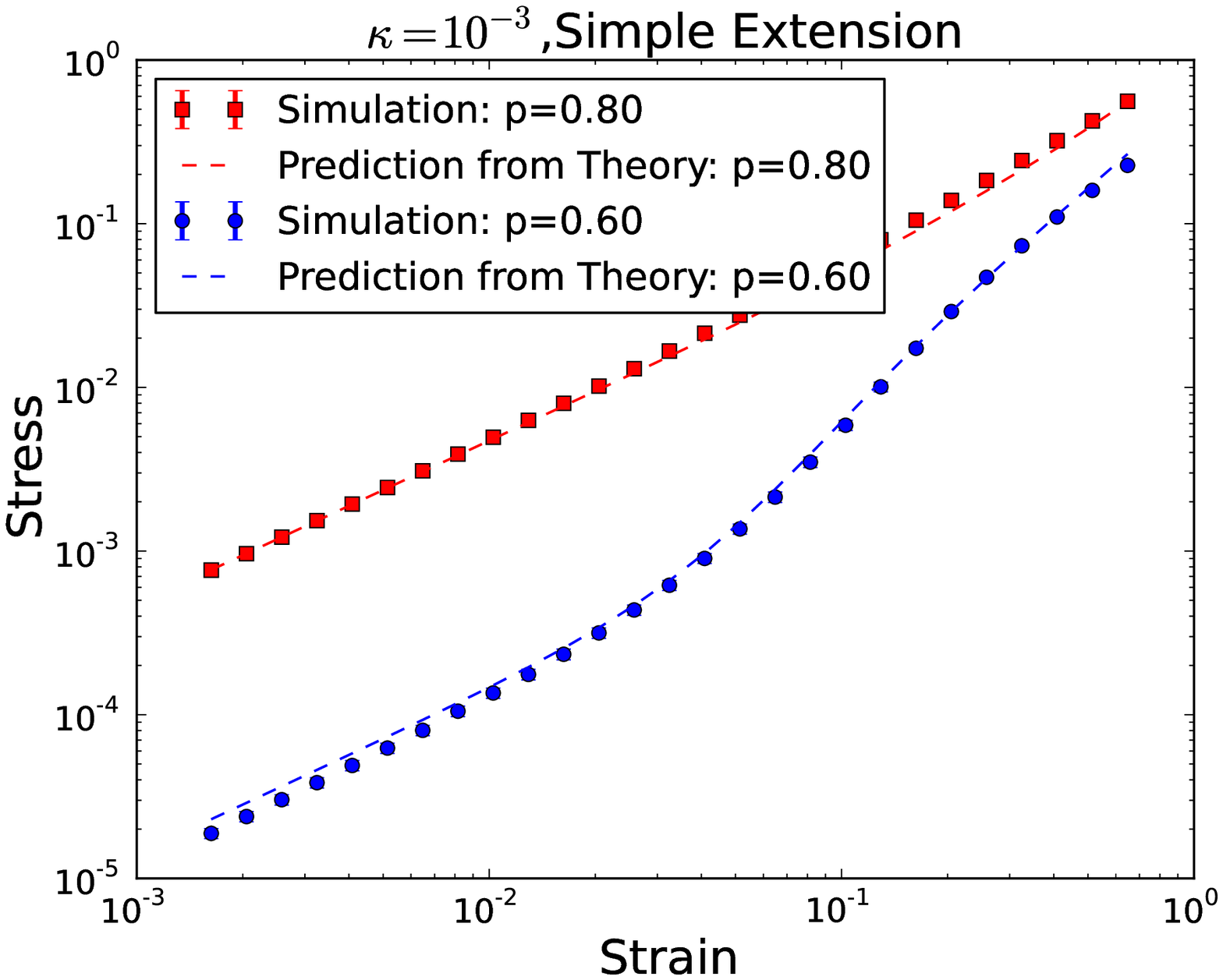}}
		\subfigure[]{\includegraphics[width=.23\textwidth]{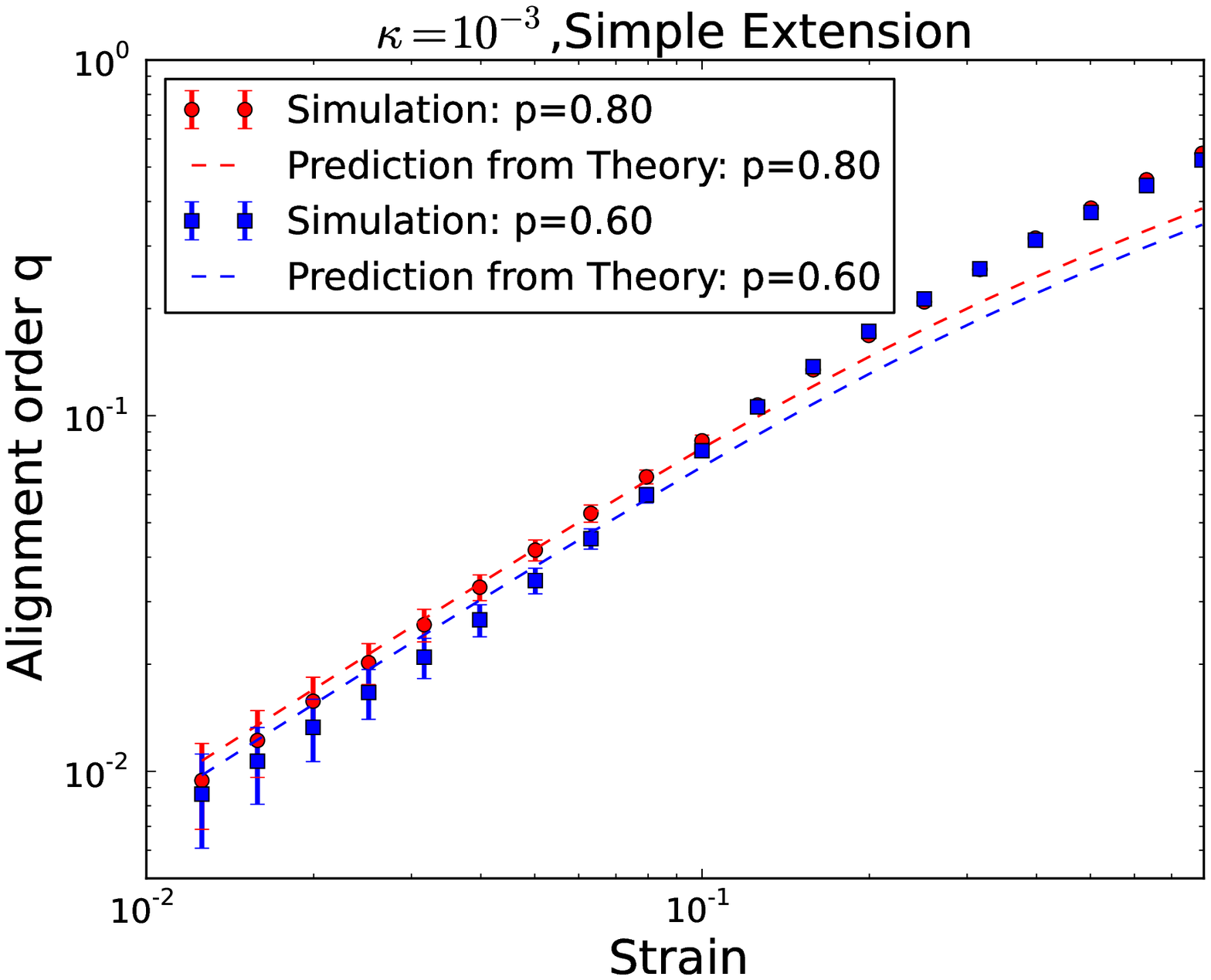}}
	\caption{Simulation results (data points) and theory  (dashed lines) for  simple extension at $p=0.6, 0.8$ using parameters fitted from shear and the hydrostatic expansion.}
	\label{FIG:Extension}
\end{figure}

\begin{figure}
	\centering
		\subfigure[]{\includegraphics[width=.23\textwidth,height=.20\textwidth]{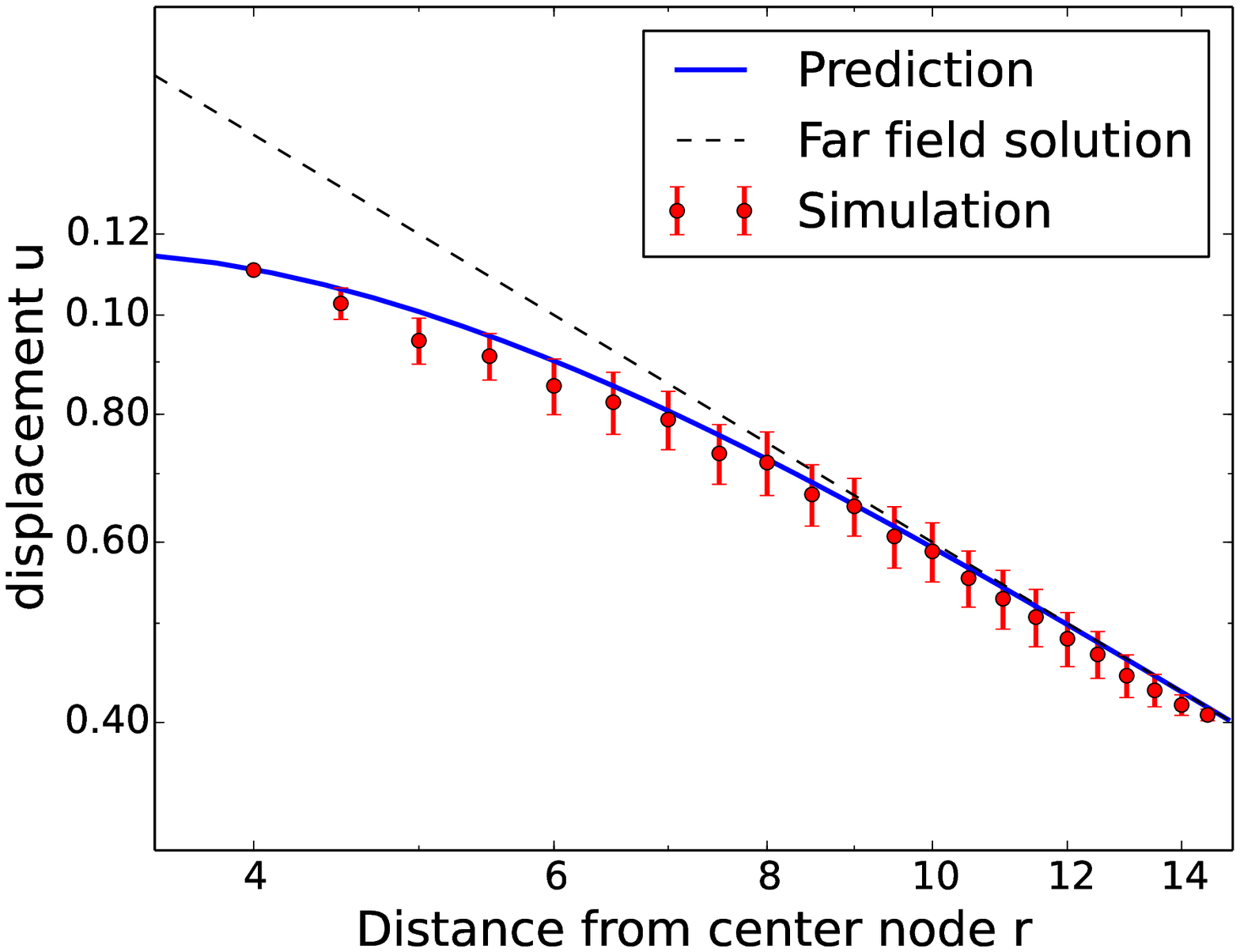}}
		\subfigure[]{\includegraphics[width=.23\textwidth,height=.20\textwidth]{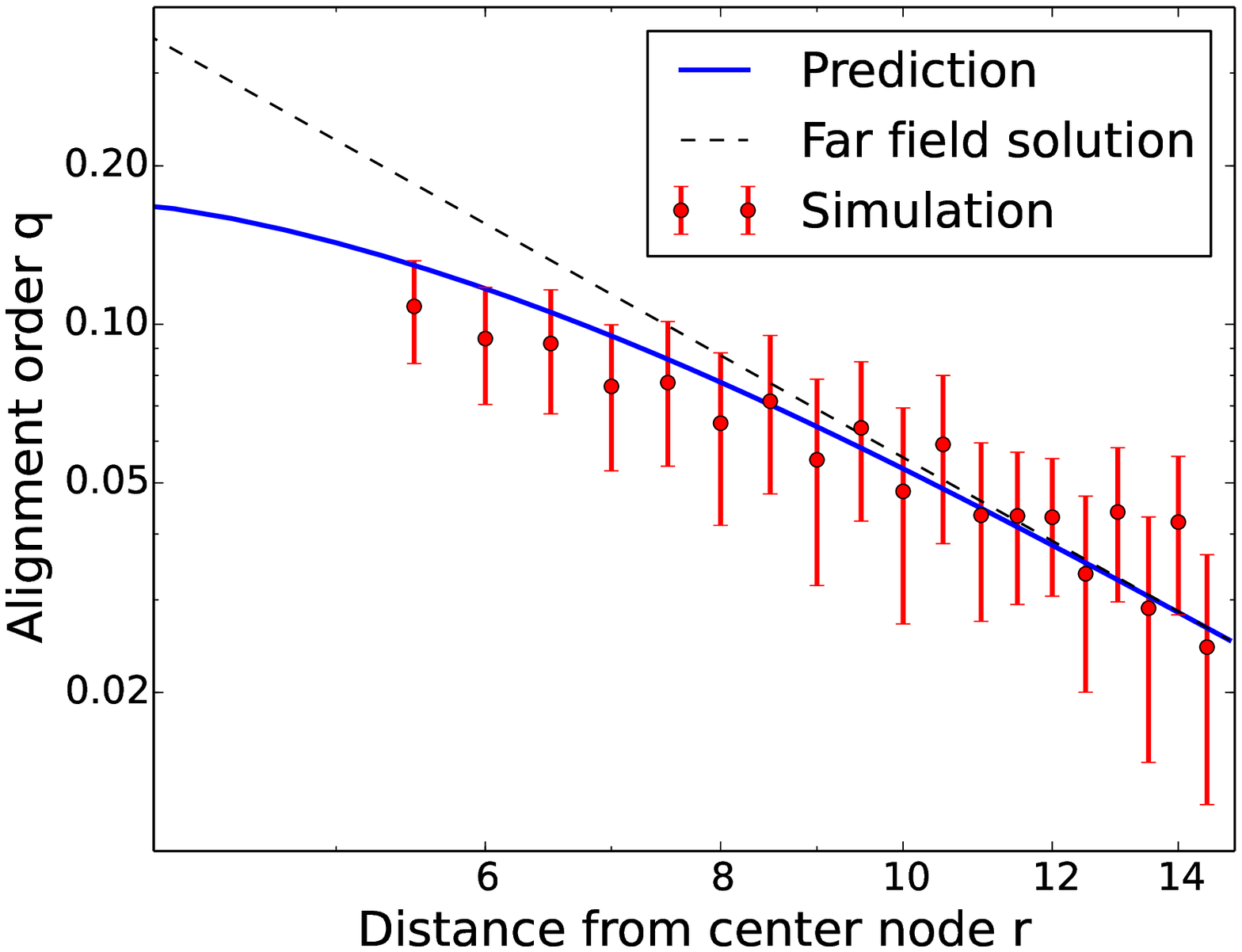}}
	\caption{Simulation results (points) and theory predictions (solid lines) for a model for a cell in ECM using the parameters from the homogeneous deformation case.}
	\label{FIG:Circular}
\end{figure}


We now put a ``cell" in the ECM by cutting a circular hole in the material  and applying a negative pressure to the exposed surface to represent the contractile force of the cell \cite{Shokef12,Sander13}.   This is  done in  Landau theory by applying boundary conditions. For the lattice model, we apply forces to nodes of the lattice at radius of several lattice spacings  and relax the lattice. A real cell is more like a force dipole, and we could use an elliptical hole to represent this. We start with the simplest case.

To use Eq.~(\ref{EQ:FE}) in this case, we note that we have two variables which depend on the distance, $r$, from the center of the hole: they are the radial component of the displacement, $u(r)$, and the  radial alignment, $q(r)$. If we minimize $F$ with respect to $u, q$ we find:
\begin{eqnarray}
\label{eq:radial}
u''(r)+\frac{u'(r)}{r}-\frac{u(r)}{r^2}&=&\frac{t}{K+\bar{\mu}}\left(\frac{q'(r)}{2}+\frac{q(r)}{r}\right)  \\
\label{eq:radial2}
\frac{t}{2}\left(u'(r)-\frac{u(r)}{r}\right)&=&V'(q).
\end{eqnarray}
For large $r$ the deformation is small, and the system is linear. If we use $V \approx Aq^2/2$ we get the familiar result for deformation in two dimensions \cite{Landau86} 
 $u(r)\sim 1/r$.   Eq. (\ref{eq:radial2}) implies $q \sim 1/r^2$. 
 As $r$ decreases we enter the nonlinear regime. Thus there is a radius within which non-linear effects and alignment are important \cite{Sander13}.  

Alignment localization of this type is captured by our theory.  For small $r$, $q$ is large and the nonlinear potential leads to a weak dependence of $q(r)$ on $r$; that is, alignment is more or less constant near the cell \cite{Sander13}. We calculated $u(r), q(r)$ by numerically solving Eq.~(\ref{eq:radial}); see Fig.~\ref{FIG:Circular}. We  did a simulation in the lattice by choosing a node and pulling in on nodes a few lattice spacings away.  The two kinds of results agree, as shown in Fig.~\ref{FIG:Circular}.   {This calculation applies equally to the case of a tumor spheroid in the ECM~\cite{Shi2014}.  
The same problem has been studied with a different elastic model in Ref.~\cite{Shokef12}, and a change of sign in strain has been reported.  We do not find this in our simulation or our model.}

Now we return to contact guidance which is experimentally studied \cite{AgudeloGarcia11,Vader09} but not well understood. Our methods give us a natural framework for modeling this effect. Suppose we have a cell moving randomly with an effective diffusion coefficient, $D_\circ$ if there is no alignment. Based on symmetry, the simplest expression for the effect of alignment is:
$\mathbf{D} = D_\circ(1+ \alpha \mathbf{Q})$, 
where $\alpha$ is a coefficient which could be measured.  {Experimental studies on three-dimensional cell migration showed interesting deviations from random walk behavior, especially anisotropy~\cite{Wu2014}, and our model may provide a theoretical framework to understand such phenomena.}

We have given a Landau theory for elastic non-linearity in biopolymer gels. The key idea is the introduction of $\mathbf{Q}$ as a part of the order parameter to measure the exhaustion of the bending modes. 
Of course, this calculation is only the beginning of a treatment for cells. 
In further study we will represent the cell as a force dipole rather than a spherical hole. And we will put more than one cell in the system to see the interactions.

\begin{acknowledgements}
We would like to thank F. MacKintosh, C. Broedersz, M. Das, and E. Ben-Jacob for useful conversations. J-C Feng is supported by the National Science Foundation Center for Theoretical Biological Physics (Grant PHY-1308264). The work of HL was supported in part by the Cancer Prevention and Research Institute of Texas
(CPRIT) Scholar Program of the State of Texas at Rice University.
\end{acknowledgements}

\end{document}